\documentclass[11pt,epsf]{article}

\usepackage{epsfig}
\usepackage{amssymb}
\usepackage{color}
\usepackage{graphicx,epsfig}

\begin{document}

\baselineskip 6mm
\renewcommand{\thefootnote}{\fnsymbol{footnote}}


\newcommand{\nc}{\newcommand}
\newcommand{\rnc}{\renewcommand}



\newcommand{\tcb}{\textcolor{blue}}
\newcommand{\tcr}{\textcolor{red}}
\newcommand{\tcg}{\textcolor{green}}


\def\be{\begin{equation}}
\def\ee{\end{equation}}
\def\ba{\begin{array}}
\def\ea{\end{array}}
\def\bea{\begin{eqnarray}}
\def\eea{\end{eqnarray}}
\def\nn{\nonumber\\}


\def\ct{\cite}
\def\la{\label}
\def\eq#1{(\ref{#1})}


\def\a{\alpha}
\def\b{\beta}
\def\g{\gamma}
\def\G{\Gamma}
\def\d{\delta}
\def\D{\Delta}
\def\e{\epsilon}
\def\et{\eta}
\def\ph{\phi}
\def\Ph{\Phi}
\def\ps{\psi}
\def\Ps{\Psi}
\def\k{\kappa}
\def\l{\lambda}
\def\L{\Lambda}
\def\m{\mu}
\def\n{\nu}
\def\th{\theta}
\def\Th{\Theta}
\def\r{\rho}
\def\s{\sigma}
\def\S{\Sigma}
\def\ta{\tau}
\def\o{\omega}
\def\O{\Omega}
\def\pr{\prime}


\def\half{\frac{1}{2}}

\def\goto{\rightarrow}

\def\na{\nabla}
\def\grad{\nabla}
\def\curl{\nabla\times}
\def\div{\nabla\cdot}
\def\pa{\partial}

\def\bra{\left\langle}
\def\ket{\right\rangle}
\def\lb{\left[}
\def\lc{\left\{}
\def\ls{\left(}
\def\lp{\left.}
\def\rp{\right.}
\def\rb{\right]}
\def\rc{\right\}}
\def\rs{\right)}
\def\fr{\frac}

\def\vac#1{\mid #1 \rangle}


\def\td#1{\tilde{#1}}
\def\check{ \maltese {\bf Check!}}


\def\Tr{{\rm Tr}\,}
\def\det{{\rm det}}


\def\bc#1{\nnindent {\bf $\bullet$ #1} \\ }
\def\ch {$<Check!>$ }
\def\ss {\vspace{1.5cm}}

\begin{titlepage}

\hfill\parbox{5cm} { }

\vspace{25mm}

\begin{center}
{\Large \bf Transport in non-conformal holographic fluids}

\vskip 1. cm
  { Shailesh Kulkarni$^a$\footnote{e-mail : skulkarnig@gmail.com},
  Bum-Hoon Lee$^{ab}$\footnote{e-mail : bhl@sogang.ac.kr},
  Jae-Hyuk Oh$^{cd}$\footnote{e-mail : jack.jaehyuk.oh@gmail.com}, 
   Chanyong Park$^a$\footnote{e-mail : cyong21@sogang.ac.kr},\\
  and Raju Roychowdhury$^a$\footnote{e-mail : raju.roychowdhury@gmail.com}
  }

\vskip 0.5cm

{\it $^a\,$Center for Quantum Spacetime (CQUeST), Sogang University, Seoul 121-742, Korea}\\
{\it $^b\,$Department of Physics, Sogang University, Seoul 121-742, Korea}\\
{\it $^c\,$Harish-Chandra Research Institute, Chhatnag Road, Jhunsi, Allahabad-211019, India} \\
{\it $^d\,$Department of Physics, Hanyang University, Seoul 133-791, Korea} \\
\end{center}

\thispagestyle{empty}

\vskip2cm


\centerline{\bf ABSTRACT} \vskip 4mm

\vspace{1cm}
We have considered non-conformal fluid dynamics whose gravity dual is a certain Einstein dilaton system with Liouville type dilaton potential, 
characterized by an intrinsic parameter $\eta$. 
We have discussed the Hawking-Page transition in this framework using hard-wall model 
and it turns out that the critical temperature of the Hawking-Page transition 
encapsulates a non-trivial dependence on $\eta$.
We also obtained transport coefficients such as AC conductivity, 
shear viscosity and diffusion constant in the hydrodynamic limit, which show non-trivial $\eta$ dependent deviations from those in conformal fluids,
although the ratio of the shear viscosity to entropy density is found to saturate the universal bound. 
Some of the retarded correlators are also computed in the high frequency limit for case study.

\vspace{2cm}


\end{titlepage}

\renewcommand{\thefootnote}{\arabic{footnote}}
\setcounter{footnote}{0}

\tableofcontents

\section{Introduction}
Gauge/gravity duality is a mighty device to study strongly interacting large N conformal field theories(CFTs) \cite{tS}-\cite{ Holography}. 
In this context, d-dimensional CFTs are mapped to gravity theories defined in asymptotically d+1 dimensional AdS spacetime, arising as its boundary theories. 
Especially, the fluid/gravity duality\cite{Janik:2005zt}-\cite{Hubeny:2011hd} has shed light on conformal fluid dynamics, which is low frequency (momentum) 
description of CFTs, and has provided much
useful information such as
transport coefficients of strongly coupled quark/gluon plasmas at RHIC and at the LHC \cite{Policastro:2001yc}.
Another interesting area is AdS/CMT \cite{Sachdev}-\cite{Liu}, which has recently witnessed a plethora of activities capturing interesting properties of various condensed matter systems as thermal phase 
transitions between super conductors and normal conductors and their critical behaviors.

While the initial interest in tangible examples in the subject concentrated on applications 
to AdS gravity theories and their CFT duals, the class of metrics of interest in gauge/gravity duality has been considerably enlarged including gravity systems, 
whose asymptotic behaviors are not AdS space time
\footnote{For 
example, there are Lifshitz geometry, non-relativistic scaling metrics characterizing the so-called 
Schrodinger spacetime and even anisotropic space as dual geometries \cite{Kachru:2008yh}-\cite{Lee:2011zzf} in the literature.}.
The dual field theories defined on the asymptotically $AdS_{d+1}$ space is conformal and enjoys $SO(2,d-1)$ conformal symmetry. However, 
it is widely accepted that one can generalize gauge/gravity duality prescription to the non-conformal field theory. Recently there is a deluge of research activities  
claiming that there are exemplar of dual field theories with gravity duals that do not possess conformal symmetries 
but another type of scaling symmetries, opulently found in Lifshitz geometry. 

From the gravity point of view, such violation of the hyperscaling symmetry \cite{Goldstein:2009cv}-\cite{Gouteraux:2011qh} can be
represented in terms of the metric whose proper distance is not invariant under an appropriate
scaling of the coordinates. It was shown in \cite{Huijse:2011ef,Ogawa:2011bz} that the theory dual to the hyperscaling violating geometry 
can have a free energy including a logarithmic term in some parameter range, which takes a typical form leading to logarithmic violation of the area law in the presence of a 
Fermi surface.  

Especially, in a recent paper \cite{Kulkarni:2012re}, the authors have considered an interesting example of non-conformal fluid dynamics whose gravity dual 
is 4-dimensional Einstein-dilaton system with Liouville type potential $V(\phi)$ for the dilaton, 
parameterized by a constant $\eta$ as $V\sim e^{\eta\phi}$, where $\phi$ is the dilaton field. 
It turns out that the dilaton solution has the form of a logarithmic function in $r$, where $r$ is the radial coordinate of
the bulk space time. This solution is asymptotically AdS when $\eta=0$. However, it shows non-trivial asymptotics of the bulk metric 
as $r\rightarrow\infty$ when $\eta$ is non-zero, presenting a deviation from an asymptotically AdS space in terms of $\eta$. 
Even if it shows such a non-trivial behavior, the boundary space time and thus its dual field theory on that are still well-defined. 
Moreover, one might expect that it may feature many interesting properties which are quite different from those in the usual conformal fluids and hence can be considered as another variant of 
fluid/gravity avatars (non-conformal fluid dynamics).

In this note, we have studied some thermodynamic properties of the non-conformal fluid dynamics and computed transport coefficients  to understand its properties more precisely. 
We have studied Hawking-Page phase transition of the system by using hard-wall model. It turns out that there is Hawking-Page phase
transition at the critical temperature 
\begin{equation}
 T^c_H=\frac{12-\eta^2}{4\pi(4+\eta^2)}2^{\frac{4-\eta^2}{12-\eta^2}}r_0^{\frac{4-\eta^2}{4+\eta^2}},
\end{equation}
which has a non-trivial $\eta$ dependence in it. $r_0$ is the location of the hard-wall and plays the role of an infrared cut off in the dual field theories. 

We have also resolved an issue concerning the curvature singularity in this background. The non-black brane geometry (black brane solution with zero size 
horizon, $r_h=0$, where $r_h$ is the horizon of the black brane) presents a curvature singularity in the interior, at $r=0$, where $\eta \neq 0$. However, in the hard wall model, it turns out that the singularity will be shielded by the hard wall and cannot be seen by an asymptotic observer. In those cases when the hard wall disappears, the black brane solution is always thermodynamically favorable than the
non-black brane counter part. Therefore, the singularity will be located inside the black brane horizon. 
In summary, the curvature singularity will not be seen at all whether there is a hard-wall or not.

We also investigated higher order corrections in $\o$ to the transport coefficients 
in hydrodynamic limit, where the hydrodynamic limit denotes that frequencies and momenta of excitations in the dual fluids
are much smaller than the inverse of mean free path, $l^{-1}_{mfp}$. $\omega$ is the frequency of the hydrodynamic modes in the dual fluid system. 
For the shear viscosity described by the bulk gravitational perturbation $h_{xy}$, we find that there is no emendation from $O(\o)$ and the result thus obtained is consistent with
the membrane paradigm result \cite{Iqbal:2008by}, where $x,y$ are boundary spatial coordinates. In addition to this, we have
obtained the leading order corrections in small $\o$ to the DC conductivity found in \cite{Kulkarni:2012re} governed by the transverse mode $A_y$, where $A_y$ is $y$-component of the bulk $U(1)$ fields. 

We have also computed the imaginary part of the Green's function using WKB approximation in
the high momenta limit and found decay behavior of the imaginary part of the two point correlation function
in the momentum space. We computed the diffusion constant from the bulk gravitational shear modes. 
The diffusion constant has a form of
\begin{equation}
\mathcal D=\frac{4+\eta^2}{12-\eta^2},
\end{equation}
while the analogous counterpart for conformal fluids defined on 3-dimensional boundary is
$\mathcal D_{conformal} = 1/3$. We also obtained the retarded Green's function from the gravitational shear modes and it shows poles at $\omega=-i\mathcal D q^2$
where $q$ is the momentum of the hydrodynamic modes\footnote{We have obtained the transport coefficients using the Kubo formula. There is another way to get the
transport coefficients by constructing the dual fluid equations of motion (e.g. See \cite{Bhattacharyya:2008jc,Banerjee:2008th,Bhattacharyya:2008ji,Oh:2010jp}). 
One can construct boundary stress-energy tensor and its conservation equations which
provide classical equations of motion of dual hydrodynamics. Some of the transport coefficients can be read off from the boundary stress tensor.}.

The outline of the paper is as follows: In section \ref{Hawking-Page transition}, we study the Hawking-Page phase transition employing hard wall model and suggest a resolution of
the curvature singularity problem in Einstein-dilaton theory.
Section \ref{transport coeffnts} is devoted to the computation of the leading order corrections in $\o$ to the transport coefficients namely shear viscosity in section
\ref{shear} and AC conductivity in section \ref{AC conduct} including a check of the universal bound on the ratio of shear viscosity to entropy density in the zero momentum limit. In section \ref{WKB analysis}, we find a decay behavior of the imaginary part
of the Green's function with momenta in high frequency and high momenta limit. Then in the following section \ref{Gravitational perturbations with odd parity in y and Retarded Green's function}, we present the crux of the gravitational perturbation with odd parity in the most general case and find the shear pole and the diffusion constant. Finally we conclude our work with some remarks in section \ref{discussion}.

\section{Hawking-Page transition: A digression}
\label{Hawking-Page transition}
\subsection{The Holographic setup}
We start with Einstein-Hilbert action with massless scalar field governed by Liouville type potential and $U(1)$ gauge fields interacting with that scalar as
\begin{equation}
\label{Lorentzian-action}
S=\frac{1}{16\pi G_4}\int dr d^3x \sqrt{-g}\left( R -2(\partial \phi)^2 -V(\phi)-\frac{1}{4g^2}e^{\alpha \phi}F_{\mu\nu}F^{\mu\nu}\right),
\end{equation}
where $V(\phi)=2\Lambda e^{\eta \phi}$, $\Lambda$ and $\eta$ are intrinsic parameters of the theory, $G_4$ is 4-dimensional {\it Planck} constant and $R$ is the curvature scalar. 
$\alpha$ is a real constant, $g$ is the gauge coupling and the 2-form field strength is given by $F_{\mu\nu}=\partial_\mu A_\nu-\partial_\mu A_\nu$, where $A_\mu$ is $U(1)$ gauge field.
We will obtain the background geometry without turning on the $U(1)$ field, i.e. $A_\mu=0$. However, we will turn on the gauge field perturbatively to compute AC conductivities in 
Sec.\ref{transport coeffnts}.
For convenience, we set $G_4=1$. 
The equations of motion are given by
\begin{eqnarray}
W_{\mu\nu}&\equiv& R_{\mu\nu}-\frac{1}{2}g_{\mu\nu}R+\frac{1}{2}g_{\mu\nu}V(\phi)-2\partial_\mu \phi \partial_\nu \phi+g_{\mu\nu} (\partial \phi)^2=0, \\ 
Y&\equiv&\frac{1}{\sqrt{-g}}\partial_\mu\left( \sqrt{-g}g^{\mu\nu}\partial_\nu \phi \right)- \frac{1}{4} \frac{\partial V(\phi)}{\partial \phi}=0.
\end{eqnarray}
The ansatz for the metric and dilaton solution are given by
\begin{eqnarray}
\label{metric-primitive}
ds^2&=&-a^2(r)f(r)d\tilde t^2+\frac{dr^2}{a^2(r)f(r)}+b^2(r)\left(d\tilde x^2+d\tilde y^2 \right), \\
\phi&=&\phi(r).
\end{eqnarray}

Exact solutions of the equations of motion are obtained as
\begin{eqnarray}
&& a(r) = \frac{\sqrt{-\Lambda} (4+\eta^2)}{2\sqrt{12-\eta^2}}e^{\frac{\eta}{2}\phi_0}r^{a_1} 
\quad ,  \quad b(r)=b_0 r^{a_1}, \nn
&& f(r)=1-\left(\frac{r}{r_h}\right)^{-c}, \\
{\rm and } \quad && \phi(r) = \phi_0-\frac{2\eta}{4+\eta^2} \ln (r),
\end{eqnarray}
where $b_0$ and $\phi_0$ are arbitrary integration constants, $r_h$ is horizon of the black brane and $a_1$ and $c$ are given by
\begin{equation}
a_1=\frac{4}{4+\eta^2} {\rm \ and \ }c=\frac{12-\eta^2}{4+\eta^2}.
\end{equation}
For later convenience, we take $b_0 = 1$, $\phi_0=0$ and
\begin{equation}
\Lambda = - \frac{4 (12 - \eta^2)}{(4 + \eta^2)^2}.
\end{equation}
Such a choice of the parameters force $a_0 =1$ and the Poincare symmetry of the boundary space time becomes manifest.
The final form of the metric and the massless scalar are given by
\begin{eqnarray}
\label{background metricTT}
ds^2&=& r^{2 a_1} \lb 1-\left({r}/{r_h} \right)^{-c} \rb d \tau^2
+\frac{dr^2}{r^{2 a_1} \lb 1-\left({r}/{r_h}\right)^{-c} \rb}+ r^{2 a_1}  \left(d x^2+dy^2 \right), \\ 
\phi&=&-\frac{2\eta}{4+\eta^2} \ln (r),
\end{eqnarray}
where $\tau$ is the Euclidean time obtained by wick rotation from the Lorentzian time $t$.
We utilize this coordinate to argue thermodynamic properties of the Einstein-dilaton system, especially to discuss Hawking-Page transition in the next subsection.

\subsection{Hawking-Page phase transition}
In this section, we would like to discuss Hawking-Page transition of the Einstein-dilaton system. In some recent papers (e.g.\cite{HP1,Gursoy:2008za}), the authors argued that there is Hawking-Page type of
thermodynamic phase transition in the dual gauge theories by finding gravitational descriptions(identifying bulk on-shell actions with free energies in
the dual field theories) where the dual gravity space time is capped off in the infrared near $r=0$. 
This is the so-called hard-wall model. 

{Notice that the radial cut off near $r=0$ in the dual gravity systems may correspond to the IR cut off in dual gauge theories.
The theory dual to the non-AdS black brane describes a (2+1)-dimensional
relativistic non-conformal theory, which would come in handy as a toy model to understand the condensed matter system. If such system has an appropriate IR cut off (the size of the
system), the dual gravity theory should also have an analogous counter part, a radial cut off.
Such a cut off can be easily realized by introducing a hard-wall.}

In the following discussion, we will restrict ourselves in the regime as $0\leq\eta^2<4$, since it was already found in \cite{Kulkarni:2012re} 
that if $\eta^2\geq4$, then the black brane solution becomes unstable.
To realize the Hawking-Page transition in this framework, we consider Euclidean version of the gravitational action of the Einstein-dilaton 
system (\ref{Lorentzian-action}) with an infrared cut off located at $r=r_0$ as
\begin{equation}
S_E=-\frac{1}{16\pi}\int^{r_{UV}}_{\tilde r}dr \int^{\tilde\beta}_0 d\tau d^2x\sqrt{g}\left( R -2(\partial \phi)^2 -V(\phi) \right),
\end{equation}
where $r_{UV}$ is the UV-cut off of the radial variable $r$ to regularize the bulk gravity action.
This is due to the fact that for both black brane and non-black brane backgrounds
(the bulk geometry (\ref{background metricTT}) with $r_h=0$), $S_E$ is infinite.
$\tilde r$ is chosen as $\tilde r=r_0$ for non-black brane background since the spacetime will be capped off at that location. For the black brane background, $\tilde r=Max(r_0,r_h)$, 
where `$Max(A,B)$' means it will pick up the bigger value between $A$ and $B$. $\tilde \beta$ is the periodicity of Euclidean time, which will be specified in a moment.

The precise integration can be performed for each of the cases.
For non-black brane solution, the regularized bulk action is given by
\begin{equation}
S^{nbb}_E=\frac{\beta V_3}{2\pi (4+\eta^2)}\left( r_{UV}^{\frac{12-\eta^2}{4+\eta^2}}- r_{0}^{\frac{12-\eta^2}{4+\eta^2}}\right),
\end{equation}
whereas the same for a black brane solution is obtained to be
\begin{equation}
S^{bb}_E=\frac{\beta_B V_3}{2\pi (4+\eta^2)}\left( r_{UV}^{\frac{12-\eta^2}{4+\eta^2}}- Max(r_{0},r_h)^{\frac{12-\eta^2}{4+\eta^2}}\right),
\end{equation}
where $\beta_B$ is periodicity of the Euclidean time for black  brane geometry, which is chosen in such a way that there is no conical singularity at the black brane horizon. 
However, $\beta$ is completely arbitrary.
We compare the two Euclidean actions at the $UV$ cut off, $r=r_{UV}$, where the local periodicity in the Euclidean time is the same, namely 
\begin{equation}
\beta=\sqrt{1-\left( \frac{r_h}{r_{UV}} \right)^{\frac{12-\eta^2}{4+\eta^2}}} \beta_B.
\end{equation}

After all this manipulation, we compute the difference between the two bulk actions $\Delta S_E$, which is given by
\begin{equation}
\Delta S_E=\lim_{r_{UV}\rightarrow \infty}\left[S^{bb}_E-S^{nbb}_E\right].
\end{equation}
The precise computation shows that 
\begin{eqnarray}
\label{exact free energy}
\Delta S_E &=& \frac{\beta_B V_3}{4\pi(4+\eta^2)}r_h^{\frac{12-\eta^2}{4+\eta^2}} {\rm \ \ \ \ \ \ \ \ \ \ \ \ \ \ \ \ \ \ \ \ for\ } r_0 > r_h, \\ \nonumber
&=&\frac{\beta_B V_3}{4\pi(4+\eta^2)} \left(  2r_0^{\frac{12-\eta^2}{4+\eta^2}} -r_h^{\frac{12-\eta^2}{4+\eta^2}}\right)  {\rm \ for\ } r_0 < r_h.
\end{eqnarray}
If $\Delta S_E$ is positive, then the non-black brane solution is thermodynamically preferable, otherwise black brane solution is preferred. Therefore, there will be a Hawking-Page type of
thermodynamic phase transition at the critical temperature
\begin{equation}
 T^c_H=\frac{12-\eta^2}{4\pi(4+\eta^2)}2^{\frac{4-\eta^2}{12-\eta^2}}r_0^{\frac{4-\eta^2}{4+\eta^2}},
\end{equation}
where we have used the expression of Hawking temperature of the Einstein dilaton system \cite{Kulkarni:2012re},
\begin{equation}
T_H=\frac{(12-\eta^2)}{4 \pi (4+\eta^2)}r_h^{\frac{4-\eta^2}{4+\eta^2}} .
\end{equation}

This result shows that for a fixed $\eta$, the critical temperature increases
when $r_0$ increases. More interesting is the behavior of the critical temperature $T^{c}_H$ which shows non-trivial $\eta$ dependence for given $r_0$. When $r_0< \frac{1}{2e^2}\sim 0.0677$, 
the critical temperature $T^c_H$ is a monotonically increasing function as $\eta$ runs from 0 to 2, whereas when $r_0 > \frac{1}{2^{1/9}e^{2/3}}\sim 0.4754$, $T^c_H$ is monotonically
decreasing. When $ \frac{1}{2e^2}<r_0 <\frac{1}{2^{1/9}e^{2/3}}$, $T^c_H$ increases upto a turning point and after passing that point, it starts to decrease. This turning point is given by
$r_0=2^{-\frac{(4+\eta^2)^2}{(\eta^2-12)^2}}e^{\frac{2(4+\eta^2)}{\eta^2-12}}$ in $r_0-\eta$ plane. Some of the intuitive examples for these curves are provided in Figure.1. 
As can be seen from Figure.1, when $\eta=2$, all the curves are converging to a single value, where critical temperature does not depend on the IR cut-off. In fact, this point is rather singular, since the black brane specific heat becomes infinite and Hawking temperature does not depend on the black brane horizon at all.
 
\begin{figure}
 \begin{center}
 \includegraphics[scale=0.9]{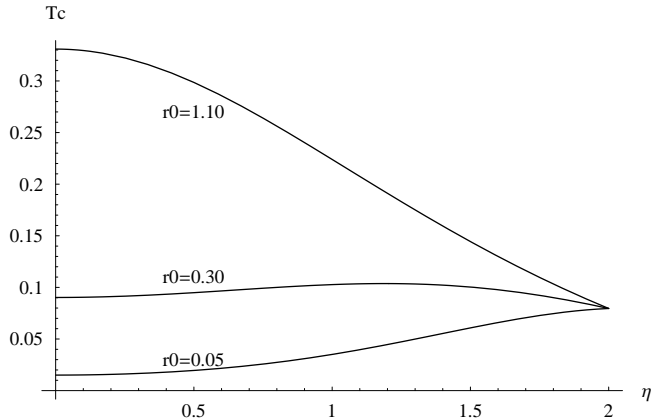}
 \caption{This graph shows the behavior of the critical temperature, $T^c_H$ with different $r_0$ values. When $r_0=1.10$, $T^c_H$ is monotonically decreasing, whereas for $r_0=0.05$, $T^c_H$ is
 monotonically increasing. If $r_0=0.30$, $T^c_H$ increases upto $\eta\eqsim1.1889$, and then starts to decrease after passing that point. All the curves converge to a single point at $\eta=2$,
 where $T^c_H=1/4\pi$.}
 \end{center}
\end{figure}

Before closing this section, we note that ``non-black brane geometry''
contains a curvature singularity at $r=0$. The curvature singularity in the interior of the ``non-black brane geometry'' is given by
\be
R =  - \fr{24 (8 - \et^2) }{(4 + \et^2)^2} \fr{1}{r^{2 \et^2 / (4 + \et^2)}},
\ee
which clearly shows a divergence unless $\eta=0$.

However, this curvature singularity doesn't pose any problem to our previous discussion regarding the Hawking-Page transition since we have ripped off the singularity from the bulk 
space time by putting an IR cut off at $r=r_0$. Even in the case of an infinitely small IR cut off $r_0$, (as seen in the second line of (\ref{exact free energy})), the black brane solution 
is always thermodynamically favorable, therefore the singularity will be concealed by black brane horizon. Moreover, it turns out that the bulk on-shell Lagrangian density
(free energy density in the dual field theory) is
finite even for a zero size IR cut off, $r_0=0$ when $0\leq\eta^2<4$. In summary, the curvature singularity is not at all problematic.

\section{Transport coefficients}
\label{transport coeffnts}
In this section, we will compute the shear viscosity, AC conductivity and diffusion constant
\footnote{We note that there are some of overlap between our results in this section and those in \cite{Springer:2008js}.
Especially, the shear viscosity and shear pole computations are performed in the same bulk background geometry and shows the consistent results with them.}.  
For these, we turn on gravitational and gauge perturbations as
\begin{equation}
g_{\mu\nu} \rightarrow { g}^{(0)}_{\mu\nu} + h_{\mu\nu}~,
\quad\quad
A_\mu  \rightarrow { A}^{(0)}_\mu + { \delta A}_\mu~.
\end{equation}
where ${g}^{(0)}_{\mu\nu}$ and ${A}^{(0)}_\mu $ are background metric and gauge fields.
${g}^{(0)}_{\mu\nu}$ is the background metric given by (\ref{background metric}) whereas ${A}^{(0)}_\mu $ is taken to be zero since there is no background gauge fields.
For the later discussion, we will drop $\delta$ from $\delta A_{\mu}$, then $A_\mu$ will denote the perturbative bulk $U(1)$ gauge fields.

In our calculations, we would like to use a new radial coordinate as 
$u \equiv \frac{r}{r_h}$ together with the rescaled variables $t$, $x$ and $y$ as
{$\{\tilde x,\tilde y\}\rightarrow \frac{1}{b_0 r^{a_1}_0} \{x,y\}$, 
$\tilde t\rightarrow \frac{1}{r_h} t$ }
and choose $\phi_0=\frac{2\eta}{4+\eta^2} \log (r_h)$(the coordinates with tilde are primitive variables in metric(\ref{metric-primitive})). 
{Furthermore, we also take
\begin{equation}
\label{lambdaaaaa}
\Lambda = - \frac{4 (12 - \eta^2)}{(4 + \eta^2)^2},
\end{equation}
as before \cite{Park:2012cu}.}
Then, the background metric and dilaton solutions become
\begin{eqnarray}
\label{background metric}
ds^2&=&-a(u)^2 f(u)d t^2+\frac{du^2}{a(u)^2 f(u)}+a(u)^2 \left(d x^2+dy^2 \right), \\ 
\phi&=&-\frac{2\eta}{4+\eta^2} \log (u),
\end{eqnarray}
where 
\begin{equation}
a(u)= u^{a_1},{\ \ }f(u)=1-u^{-c} .
\end{equation}
In the following discussions of the transport coefficients, we will mostly use the rescaled coordinate. 
In this coordinate, the horizon is located at $u=1$. The black brane temperature is given by
{\begin{equation}
T_H 
= \frac{c}{4 \pi},
\end{equation}}
and entropy of the black brane is $S=\frac{V_2}{4}$, where $V_2= \int dx dy$ is the spatial volume of the boundary space time.

To obtain hydrodynamic modes in the dual fluids, we turn on metric perturbations together with dilaton perturbations. For a better computation, 
we classify the gravitational and dilaton modes as follows. The background metric solution(\ref{background metric}) 
enjoys $SO(2)$ global symmetry (spatial rotation in $x-y$ plane) together with
parity  symmetries as $x\rightarrow -x$ and $y\rightarrow -y$. Therefore, perturbations can be
classified in accord with these global symmetries when they depend on time only. They are given by Table.1

\begin{table}[tp]%
\label{table1}
\caption{Modes Classifications of time dependent perturbations.}
\label{{aggiungi}}\centering%
\begin{tabular}{clccc}
 \hline\hline
Gauge Fields& Each Decoupled Mode&Global $SO(2)$& Parity in x& Parity in y\\
\hline
NIL& $h_{xy}$& Tensor & Odd & Odd\\
NIL& $h_{xx}-h_{yy}$ & Tensor & Even & Even\\
$A_{x}$& $h_{tx}$ & Vector & Odd & Even\\
$A_{y}$& $h_{ty}$ & Vector & Even & Odd\\
$A_{t}$& $h_{tt}$, $h_{xx}+h_{yy}$ and $\phi$ & Scalar & Even & Even\\
\hline
\end{tabular}
\end{table}

However, once one turns on spatial momenta, it breaks these symmetries into smaller symmetry group. Without any loss of generality, we can choose the direction of the spatial momenta along  
$x$-direction. In this case, global $SO(2)$ symmetry breaks into $U(1)$ and only $y \rightarrow -y$ parity symmetry is retained. Global $U(1)$ is just a translational symmetry along $y$ 
direction and is not relevant for the mode classification. Therefore, we sort out the perturbative modes by $y$ parity symmetry. They are given by Table.2.

For the shear viscosity one considers a tensor mode of the metric fluctuation denoted by $h_{xy}$. 
For the computation of electrical conductivity using Kubo's formula, it is sufficient to consider gauge fluctuations 
around the black brane solution and keep other fields unperturbed. 
In order to determine diffusion constant, it is enough to  consider perturbations in $(ty)$ and 
$(xy)$ component of the metric tensor. 

\begin{table}[tp]%
\label{table2}
\caption{Modes Classifications of frequency and momenta dependent perturbations.}
\label{{aggiungi}}\centering%
\begin{tabular}{clccc}
 \hline\hline
Gauge Fields& Each Decoupled Mode& Parity in y\\
\hline
$A_{y}$& $h_{ty}$ and $h_{xy}$& Odd\\
$A_{t}$ and $A_{x}$& $h_{tt}$, $h_{xx}$, $h_{yy}$, $h_{tx}$ and $\phi$ & Even \\
\hline
\end{tabular}
\end{table}

\subsection{The shear viscosity and AC conductivity in the zero momentum limit}

In the zero momentum limit, the gauge fluctuations are decoupled from the metric ones in
Table 2. In this limit, the study of the transverse gauge fluctuation with nonzero frequency
provides information about the AC conductivity, which reduces to the DC conductivity in the 
zero frequency limit \cite{Lee:2010qs}. In addition, we can find the shear viscosity from the bulk gravitational perturbation $h_{xy}$.

\subsubsection{The shear viscosity} 
\label{shear}

Consider a tensor mode of the metric fluctuations, $h_{xy}$.
In the zero momentum limit, using the following Fourier transform as
\be
{h^x_y} (u,t) = \int \frac{dt}{2 \pi} e^{- i \o t}  {h^x_y} (u,\o) ,
\ee
Einstein equations for $h_{xy}$ at the linear order in weak field expansion is given by
\be			\la{eq:tensormode1}
0 = {h^x_y}'' + \ls \frac{f'}{f} +\fr{4 a_1}{u} \rs  {h^x_y} ' + 
\frac{\o^2}{u^{4 a_1} f^2}  {h^x_y} ,
\ee
where the prime denotes the derivative with respect to $u$ and ${h^x_y} = {h^x_y} (u, \o)$.
The solution of the above equation can be expanded as
\be
\label{an:shearvisc}
h^x_y = f^{\d} \ls G_0 (u) + \o G_1 (u) +  \o^2 G_2 (u)  \cdots \rs,
\ee  
where $\delta$ is a constant.
Notice that the horizon lies at $u=1$ and $f = 0$ at the horizon. Due to the singular structure
of \eq{eq:tensormode1} at the horizon, the near horizon behavior of $h^x_y$ is given by
\be
h^x_{y} = f^{ \pm i \o/c} ,
\ee
where $f^{- i \o/c}$($f^{+ i \o/c}$) implies the incoming (outgoing) solution.
In order to study the transport coefficients from boundary retarded Green's functions, we should choose an incoming solution.
We solve the equation (\ref{eq:tensormode1}) order by order in small frequency and the zeroth order equation in $\o$ is
\be		\la{eq:tensormode}
0 = \ls  u^{4 a_1} f  G_0 (u)' \rs' .
\ee
The regular solution(at the horizon) of the equation is given by $G_0(u) = c_0$ where $c_0$ is an arbitrary 
constant. 
The first order equation in $\o$ is of the same form as (\ref{eq:tensormode}) and thus its solution is given by
\be
G_1 (u) = c_1 + c_2 \fr{16 \log u - (4 + \et^2) \log \ls
u - u^{16/(4 + \et^2)} \rs }{12 - \et^2} .
\ee
The regularity at the black brane horizon forces $c_2=0$. On AdS boundary, we impose (the so-called) ``vanishing condition'' as 
\begin{equation}
\label{dbc}
h^x_y(u\rightarrow 1)=c_0,
\end{equation}
then we set $c_1=0$ under such boundary condition\footnote{Under the vanishing condition, the boundary value of $h^x_y$(at $u=\infty$) is not precisely $c_0$, but get some corrections.}.

It may be interesting to ask how the form of the shear viscosity gets modified after turning on higher order 
corrections in small frequency. Such corrections can be obtained if we take a parametric regime where frequency is small but finite.
Since the first order correction in $\o$ vanishes, $G_1(u)=0$, in order to study non-trivial small frequency corrections, we solve the second order equation in $\o$
in \eq{eq:tensormode1}. The second order equation in $\o$ provides a solution
$G_2 (u)$, which turned out to be of an integral form as 
\be		\la{sol:shear}
G_2 (u) = - \int^u_1
 \fr{ d u^\prime}{u^{\prime4 a_1} f(u^\prime)} \left(\int^{u^\prime}_1 d u^{\prime\prime} 
\fr{c_0(1 - u^{\prime\prime\frac{16}{4 + \et^2}})}{u^{\prime\prime} - u^{\prime\prime\frac{16}{4 + \et^2}}} 
+ c_3\right)  +c_4 ,
\ee
where $c_3$ and $c_4$ are integration constants. $c_3$ is determined from regularity condition at the horizon and it turns out that $c_3=0$. 
The vanishing condition at the black brane horizon (\ref{dbc}) forces $G_2(u\rightarrow 1)=0$. This condition gives $ c_4=0$.
In fact, analytic form of $G^\prime_2(u)$ can be obtained as
\begin{equation}		\la{sol:shear}
G^\prime_2 (u) =0
-\fr{c_0}{u^{4 a_1} f(u)} \left.\left\{ u^{\prime}_2H_1(1,-\frac{4+\eta^2}{12-\eta^2};1-\frac{4+\eta^2}{12-\eta^2};u^{\prime-\frac{12-\eta^2}{4+\eta^2}})
 -\frac{4+\eta^2}{12-\eta^2}ln(1-u^{\prime-\frac{12-\eta^2}{4+\eta^2}})\right\}\right|^u_{1},
\end{equation}
where $\ _2H_1$ is Hypergeometric function.

Using all the above information, we evaluate bulk action upto equations of motion for $h^x_y$, which is given by
\begin{eqnarray}
\label{hxy-action}
S &=& - \fr{1}{32 \pi}   \int d^3 x  \ u^{4 a_1} f(u) h^x_y(u) \ {{h^x_y}'}(u) +S_{ct}\\ \nonumber
&=&-\frac{1}{32\pi}\int d\omega d^2q \tilde c_{0,\omega}\tilde c_{0,-\o}\left[ i\o -\o^2 \frac{4+\eta^2}{12-\eta^2}\left( -\gamma - PG\left(-\frac{4+\eta^2}{12-\eta^2}\right)\right)+O(\o^3)\right],
\end{eqnarray}
where $\tilde c_0$ is the boundary value of $h^x_y$(at $u=\infty$), $\gamma$ is Euler number, $PG$ is Polygamma function and $S_{ct}$ is counter term action
\footnote{The boundary value of $h^x_y$ is given by
\begin{equation}
\tilde c_0= c_0- \o^2 \int^\infty_1
 \fr{ d u^\prime}{u^{\prime4 a_1} f(u^\prime)} \left(\int^{u^\prime}_1 d u^{\prime\prime} 
\fr{c_0(1 - u^{\prime\prime\frac{16}{4 + \et^2}})}{u^{\prime\prime} - u^{\prime\prime\frac{16}{4 + \et^2}}} 
\right).
\end{equation}
}. As long as $0 \leq \eta^2 <4$, $PG\left(-\frac{4+\eta^2}{12-\eta^2}\right)$ is purely real. 
The retarded Green's function can be read off from (\ref{hxy-action}) as
\begin{equation}
G^R_{xy,xy}=\frac{1}{16\pi}\left[ i\o -\o^2 \frac{4+\eta^2}{12-\eta^2}\left( -\gamma - PG\left(-\frac{4+\eta^2}{12-\eta^2}\right)\right)\right],
\end{equation}
which shows corrections in $O(\o^2)$.
%
We define shear viscosity function as
\be
\tilde\et(\o) \equiv  \fr{Im[G^R_{xy,xy}(\o)] }{\o}  ,
\ee
and in the $\o=0$ limit it is identical to the shear viscosity obtained from the Kubo formula. We evaluate
$\tilde\et(\o)$ in the hydrodynamic limit and it may show sub-leading corrections in $\omega$ to the shear viscosity obtained from the Kubo formula.
However, it turns out that it does not contain higher order corrections in small frequency upto $O(\o)$, since the shear viscosity function 
cares only about the imaginary parts of the retarded Green's function. Therefore,
\be
\tilde\et (\o) = \fr{1 }{16 \pi} + O(\o^2).
\ee
The shear viscosity saturates the universal bound upto the first sub-leading order correction in $\o$,
$\fr{\et}{s} = \fr{1}{4 \pi}$, being consistent with the result obtained from the
membrane paradigm \cite{Iqbal:2008by,Park:2012cu}, where $s$ is the entropy density in the rescaled coordinate, $s=\frac{1}{4}$. In addition, our study on the higher
order corrections to the shear viscosity shows $\frac{\pa \et(\o)}{  \pa \o}=0$ in the zero
frequency limit.

\subsubsection{AC conductivity}
\label{AC conduct}

In \cite{Kulkarni:2012re,Park:2012cu}, DC conductivities of the non-conformal holographic fluids was investigated. 
In this section, we will study
AC conductivity of the fluids by using its gravity dual, Einstein-dilaton system.

The gauge fluctuation coupled with dilaton field is given by
\be			\la{act:gaugefl}
S_A = - \fr{1}{4 g^2} \int d^4 x \sqrt{-g} e^{\a \ph} F^{\m \n} F_{\m \n} ,
\ee
where $g^2$ is the gauge coupling.
The equations of motion of transverse modes,
$A_x$ and $A_y$ in the zero momentum limit (See Table.1), is given by
\be
0 = A'' + \ls \fr{b}{u} + \fr{f'}{f} \rs A' + \fr{\o^2}{u^{4 a_1} f^2} A ,
\ee
where $A$ indicates either $A_x$ or $A_y$ and 
\be
b = \fr{2 (4 - \a \et)}{4+\et^2} .
\ee
Notice that in the zero momentum limit, two transverse modes $A_x$ and $A_y$ satisfy the 
same differential equation due to the global SO(2) symmetry(rotation in $x-y$ plane). 

The near horizon solution of $A$ has a form 
\be
A \sim f^{\pm i \o/c} ,
\ee
in which the positive (negative) sign is for the outgoing (incoming) boundary condition
at the horizon. Imposing the incoming boundary condition on $A$, the
solution in the hydrodynamic regime ($\o<<1$) can be expanded in small frequency as
\be
A = f^{- i \o/c} \ls G_0 (u) + \o G_1 (u) + \o^2 G_2 (u) + \cdots \rs ,
\ee
in which the ellipsis indicates the higher order corrections in $\o$.

The zeroth order equation in $\o$ is given by
\be
0 = u f G_0'' + (b f + u f') G_0' ,
\ee
and its solution is obtained as
\be
G_0 = c_1 + \fr{ c_2 }{1 - b + c} \ u^{1-b+c} 
{_2 F_1} \ls \fr{1-b+c}{c},1 ,1+ \fr{1-b+c}{c},u^c \rs .
\ee
The regularity on $G_0$ at the horizon forces $c_2$ to be zero, then we get 
\be
G_0 = c_1 .
\ee

The first order equation in $\o$ is   
\be
0 = G_1 '' + \fr{c + b (u^c - 1)}{u (u^c - 1)} G_1' + \fr{i (1 - b +c)}{u^2 (u^c - 1)} G_0 ,
\ee
This equation has the following solution:	
\be
G_1 = c_3 + \fr{c_4}{1-b+c} \ u^{1-b+c}  {_2 F_1} \ls \fr{1-b+c}{c},1 ,1+ \fr{1-b+c}{c},u^c \rs
+ \fr{i c_1}{c} \log \fr{u^c - 1}{u^c} .
\ee
At the horizon, the regularity fixes $c_4$ to be
\be
c_4 = i c_1
\ee
and the vanishing condition of $G_1$ determines $c_3$ as
\be
c_3 = \fr{i c_1 \lb EG + i \pi + PG \ls 0,\fr{1-b+c}{c} \rs \rb }{c} ,
\ee
where $EG$ and $PG$ imply the Eulergamma and Polygamma functions respectively.

Finally, the second order equation in $\o$ is obtained as
\be
0 = \pa_u \ls u^b f G_2' \rs + \fr{c_1}{f} u^{-2 - 4a_1 +b -2 c} (u^{2 + 2 c} - u^{4 a_1})
+ i u^{4 a_1 + c} \lb (1-b+c) G_1 - 2 u G_1' \rb .
\ee
Its solution has an integral form like
\be			\la{sol:intform}
G_2 = - \int^u_1 du \ \fr{ I_1 + I_2 + c_5 }{u^b f}  + c_6 ,
\ee
where
\bea
I_1 &=&  c_1 \lb \fr{{_2 F_1} \ls - \fr{1- b +c}{c},1 , - \fr{1-b}{c},u^c \rs}{(1-b+c) \ u^{1 - b+c}} 
+  \fr{u^{1 - 4 a_1 + b + c } \ {_2 F_1} \ls\fr{1 - 4 a1 + b + c }{c},1 , \fr{ 1 - 4 a1 + b + 2c}{c},u^c \rs}{ 1 - 4 a1 + b + c }  \rb  , \nn
I_2 &=& c_1 \lb \fr{(3 + 8 a_1 - b + 3 c) \  u^{1 + 4 a1 + c}
 \ _2 F_1 \ls 1, \fr{1 + 4 a1 + c}{c}, 2 + \fr{1 + 4 a1}{c}, u^c \rs}{ (1 + 4 a_1 + c)^2 } \rp \nn
&& \qquad - \ \fr{2 \ u^{2 + 4 a_1 - b + 2 c} \ _2 F_1 \ls \fr{2 + 4 a_1 - b + 2 c}{c}, 1, 
\fr{2 + 4 a_1 - b + 3 c}{c}, u^c \rs}{2 + 4 a_1 - b + 2 c}  \nn
&& \qquad + \ \fr {u^{2 + 4 a_1 - b + 2 c}
  \ _2 F_1 \ls 1, \fr{1 - b + c}{c}, 2 + \fr{1 - b}{c}, u^c \rs}{1 + 4 a_1 +c} 
  - \ \fr{(1 - b + c)}{c\  (1 + 4 a_1 + c)}{\cal B} 
\ls u^c, 2 + \fr{2 + 4 a_1 - b}{c}, 0 \rs \nn 
&& \qquad \lp +  \ \fr{(1 - b + c) u^{1 + 4 a_1 + c} }{c\  (1 + 4 a_1 + c)} \lc 
  i \pi +  {\cal HN} \ls \fr{1 - b}{c} \rs  
   + \log \fr{ u^c -1}{u^c}  \rc  \rb .
\eea
${\cal HN}$ and ${\cal B}$ denote the harmonic number and beta function respectively.
$c_5$ and $c_6$ are integration constants, which can be
determined in terms of $c_1$. To determine them, after performing the integration in (\ref{sol:intform}) near the horizon,
we impose the regularity and vanishing condition. The regularity at the horizon fixes
$c_5$ to be 
\bea
c_5 &=&  c_1 \lb \fr{(3 + 8 a_1 - b + 3 c)}{c \ (1 + 4 a_1 + c)}    
\lc PG \ls 0, \fr{1 + 4 a_1 + c}{c} \rs - \ PG \ls 0, 2 + \fr{2 + 4 a_1 - b}{c} \rs \rc  \rp \nn
&& \quad \lp + \fr{1}{c} \lc PG \ls 0, \fr{1 - 4 a_1 + b + c}{c} \rs -  PG \ls 0, -1 + \fr{-1 + b}{c} \rs  \rc \rb ,
\eea
where 
the vanishing condition sets  $c_6=0$ because the integration in \eq{sol:intform}
does not give any constant at the horizon. It is worth noting that the fact $c_6 = 0$ implies that
the $\o^2$ order solution does not contribute to the boundary value of $A$.

Imposing the Dirichlet boundary condition at the asymptotic boundary like
\be			\la{cond:dba}
A_0 = \lim_{u \to \infty} A(u) .
\ee
$c_1$ can be determined by
\be
c_1 = \fr{c A_0}{c- \pi \o  + i \o  \ \lb EG + PG \ls 0,\fr{1-b+c}{c} \rs \rb} .
\ee
Bulk action \eq{act:gaugefl}
together with \eq{cond:dba} and bulk equations of motion becomes
\be		\la{act:bda}
S_A = - \lim_{u \to \infty} \fr{1}{g^2} \int d^3 x \ A_0 \ u^b f(u) \ A(u)'.
\ee
One can see that the terms proportional to $u^{-b}$ in $A(u)'$ can only contribute
to the finite part of the retarded Green's function of the transverse mode. 
Using this fact, the imaginary part of the retarded Green's function reads as
\be
G_{xx} = \fr{i \o}{g^2} \ls 1 + \fr{\pi}{c}  \ \o \rs .
\ee
After all dusts get settled, the AC conductivity becomes
\be
\s_{AC} \equiv \fr{G_{xx}}{i \o}= \fr{1}{g^2} \ls 1 + \fr{(4 + \et^2 ) \ \pi }{12-\et^2}  \ \o \rs .
\ee
In the zero frequency limit ($ \o \to 0$), the AC conductivity reduces to the DC conductivity,
which is consistent with the result obtained in \cite{Kulkarni:2012re} and also in \cite{Park:2012cu} by the membrane paradigm. By using $u =r/ r_h$ and $\o = r_h \td{\o}$, we get
\be
\s_{AC} 
=  \fr{1}{g^2} \ls \fr{ 4 \pi (4 + \et^2 )  }{12-\et^2} 
\rs^{- \fr{2 \a \et}{4 - \et^2}}  T_H^{- \fr{2 \a \et}{4 - \et^2}} + 
\fr{\td{\o}}{4 g^2}  \ls \fr{ 4 \pi (4 + \et^2 )  }{12-\et^2} 
\rs^{\fr{8-2 \a \et}{4 - \et^2}}  T_H^{\fr{4 - 2 \a \et + \et^2}{4 - \et^2}}  
+ {\cal O} (\td{\o}^2),
\ee
{where the mass dimension of $\td{\o}$ is the same as that of $1/r_h$}.
In the hydrodynamic limit, the conductivity increases with increasing frequency.
The conductivity increases (or decreases) with temperature for $\a < 0$ (or $\a > 0$), 
which shows that for $\a>0$ the dual field theory has a metal-like behavior 
(DC conductivity decreases with increasing temperature), whereas for $\a < 0$ it shows 
electrolyte-like behavior
with a decreasing DC conductivity with temperature.
Especially, the known metallic behavior for the resistivity occurs at $\a = \fr{4 - \et^2}{2 \et}$,
which shows the resistivity to be linearly proportional to the temperature $\r \sim T_H$.   

\subsection{Green's function at high momenta}
\label{WKB analysis}
In this subsection, we will consider the high momenta ($\o,|q| >>1$) domain of gauge  perturbations discussed in the previous subsections. We use a similar technique 
as given in \cite{Policastro:2002se}. We first  turn on the momentum $q$ along $x$ direction. Then the gauge fluctuations $A_{t}$ and $A_{x}$ are coupled while $A_{y}$ represents the transverse mode. This transverse mode satisfies
 \be
0 = A''_{y} + \ls \fr{b}{u} + \fr{f'}{f} \rs A'_{y} + \fr{\o^2 - f q^{2}}{u^{4 a_1} f^2} A_{y} , \la{eq: Ay}
\ee
Now, we make the transformation 
\be
A_{y}(u) = \ls\frac{u^{c-b}}{u^{c}-1}\rs^{1/2} \bar{A}_{y}(u) .
\ee
to bring (\ref{eq: Ay}) in the Schr\"{o}dinger form 
\be
 \bar{A}''_{y}(u) = \lb \o^2 Q(u) + R(u) \rb  \bar{A}_{y}(u), \la{eq: barAy} 
\ee
where 
\be
Q(u) = \frac{p^{2}f - 1}{u^{4a_{1}}f^{2}}   
\ee
and
\be
R(u) = \frac{1}{4u^2} \lb \frac{c^2}{(u^c -1)^2} + \frac{2c(c-b+1)}{u^c -1} - b(b-2)\rb
\ee
with $p = |q|/ \o$.

In the large frequency limit ($\o >> 1$), the term proportional to $\o^2$ dominate the potential and the WKB approximation is possible. To apply the WKB analysis, we first expand $ \bar{A}_{y}(u)$ as
\be
 \bar{A}_{y}(u) = \exp\lb \o\sum_{i=0}^{\infty} \o^{-i} S_{i}\rb
\ee
After substituting this WKB ansatz in (\ref{eq: barAy}) we get, at zeroth order in  $1/ \o$ 
\bea
S_{0} &=& \pm \int \sqrt{Q(u)}  \la{eq: zeroorder} 
\eea
For the spacelike momenta, $\o^2 < |q|^2 $ ($p>1$), the potential $Q(u)$ is positive in the range $u_{0} = (1-\frac{1}{p^2})^{-1/c} \le u \le \infty$ while it is negative otherwise. Hence, the WKB solution for (\ref{eq: barAy}) decays exponentially in the interval ($u_{0}, \infty$) and oscillates in the interval ($1, u_{0}$). Physically, this implies that the incoming wave first tunnel from $u=\infty$ to $u=u_{0}$ before it reaches the horizon $u=1$. The imaginary part of the retarded Green's function $G^{R}_{yy}$, which is proportional to the tunneling probability, is then given by
\be
\Im G^{R}_{yy}(\o, q)  \sim \exp  \lb  -2\o \int_{u_{0}}^{\infty} \sqrt{Q(u)} \rb \la{eq: Gyy1}
\ee
and in the limit where $\o \ll |q|$, it becomes
\be
\Im G^{R}_{yy}(\o, q)  \sim 
\exp  \lb  -2\o p \int_{1}^{\infty} \frac{u^{-2a_{1}}}{\sqrt{1-u^{-c}}} \rb \sim \exp \left(-\frac{\lambda|q|}{T_{H}} \right),  
\la{eq: Gyy2}
\ee
where
\begin{equation}
 \lambda=\lb \frac{2\sqrt{\pi}\Gamma[(2a_{1}-1)/c]}{c\Gamma[(4a_{1}-2+c)/2c]}\rb.
\end{equation}
Note that, the above expression for Green's function is independent of dilaton parameter $\alpha$. In other words, in the regime of high momenta the  dilaton field does not affect the 
retarded Green's function. This decaying nature of the imaginary part of the Green's function is similar to the one obtained in \cite{Policastro:2002se}. However, in our case the decaying 
probability depends upon the parameter $\eta$. The constant $\lambda$ is a positive number, bounded from below by its
minimum value as $\lambda\geq\lambda_{min} =\frac{2\sqrt{\pi}\Gamma(1/3)}{3\Gamma(5/6)}\sim 2.8044$ in the region $0<\eta^2<4$, which ensures
a decay behavior of the Green's function in such window for $\eta$. 

\subsection{Gravitational perturbations with odd parity in $y$ and Retarded Green's function}
\label{Gravitational perturbations with odd parity in y and Retarded Green's function}
In this section, we will turn on the parity odd modes, $h_{ty}$ and $h_{xy}$ and obtain shear modes in the dual fluid dynamics.
Non trivial equations of motion up to first order in weak field expansion are given by
\begin{eqnarray}
W_{ry}&=&\omega X^{\prime}_1(u)+q f(u)X^\prime_2(u)=0, \\
\label{ty-eq}
W_{ty}&=&X^{\prime\prime}_1(u)+\frac{4a^\prime(u)}{a(u)}X^\prime_1(u)-\frac{q^2X_1(u)}{a^4(u)f(u)}-\frac{q\omega X_2(u)}{a^4(u)f(u)}=0, \\
W_{xy}&=&X^{\prime\prime}_2(u)+X^\prime_2(u)\left(4\frac{a^\prime(u)}{a(u)}+\frac{f^\prime(u)}{f(u)} \right)+\frac{\omega^2 X_2(u)}{a^4(u)f^2(u)} 
+\frac{q\omega X_1(u)}{a^4(u)f^2(u)}=0,
\end{eqnarray}
where the perturbative fields are obtained in the frequency(momentum) space by the Fourier transform defined in the previous section
and the fields $X_1(u)$ and $X_{2}(u)$ are given by
\begin{equation}
h^{\omega,q}_{ty}\equiv a^2(u)X_1(u) {\rm \ and\ }h^{\omega,q}_{xy} \equiv a^2(u)X_{2}.
\end{equation}
$X_1$ and $X_2$ definitely have frequency and momentum dependence but we suppress the explicit expression of those for the sake of calculation.
To solve the above set of coupled differential equations, we rewrite them only in terms of $X_1$ by combining them (with the aid of a little algebra) \cite{Policastro:2002se,Alex1} and the final equation to solve is given by
\begin{equation}
0= \left( f(u)\left(  a^4(u) X^\prime_1(u)\right)^{\prime} \right)^{\prime}-\left(q^2-\frac{\omega^2}{f(u)}\right)X^\prime_1(u).
\end{equation}
This is a third order equation and we try the following ansatz for the solution:
\begin{equation}
Y_1(u)\equiv X^\prime_1(u)=C\left(1-u^{-\frac{12-\eta^2}{4+\eta^2}}\right)^{-\frac{4i\omega}{(-\Lambda)(4+\eta^2)}}G(u),
\end{equation}
where
\begin{equation}
G(u)=\sum_{i,j=0}^{\infty}\omega^i q^j Y_{1,ij}(u),
\end{equation}
$C$ is an overall constant and the factor in front of $G(u)$ is for the ingoing boundary condition near black brane horizon. It turns out that each of $Y_{1,ij}$ are given by
\begin{eqnarray}
Y_{1,00}&=&u^{-\frac{16}{4+\eta^2}} \\
Y_{1,10}&=&C^1_{10}u^{-\frac{16}{4+\eta^2}}+\frac{4i(\eta^2-12)}{-\Lambda (4+\eta^2)^2}u^{-\frac{16}{4+\eta^2}}\int^u_1
du\left(\frac{1-u^{-\frac{12-\eta^2}{4+\eta^2}-1}}{1-u^{-\frac{12-\eta^2}{4+\eta^2}}}\right)\\ \nonumber
&=&C^1_{10}u^{-\frac{16}{4+\eta^2}}+\frac{4i(\eta^2-12)}{-\Lambda (4+\eta^2)^2}u^{-\frac{16}{4+\eta^2}}
\left(u_2H_1(1,-\frac{4+\eta^2}{12-\eta^2};1-\frac{4+\eta^2}{12-\eta^2};u^{-\frac{12-\eta^2}{4+\eta^2}})\right. \\ \nonumber
&-&\left.\frac{4+\eta^2}{12-\eta^2}ln(1-u^{-\frac{12-\eta^2}{4+\eta^2}})\right) \\
Y_{1,01}&=&0,{\ \ }Y_{1,02}=u^{-\frac{16}{4+\eta^2}}\left( C^1_{02} +\frac{4u}{-\Lambda(4+\eta^2)}\right), {\rm\ \ and\ so \ on,}
\end{eqnarray}
where $C^1_{ij}$ are $O(1)$ constants and $\ _2H_1$ is Hyper geometric function.
To obtain shear pole (diffusion constant), we plug the solutions into (\ref{ty-eq}). With boundary values of the fields, $X_1(u)$ and $X_2(u)$ as
\begin{equation}
X_1(u\rightarrow \infty)\equiv \bar X_{1} {\rm\ \ and\ \ }X_2(u\rightarrow \infty)\equiv \bar X_{2},
\end{equation}
we obtained the overall coefficient $C$ in terms of those as
\begin{equation}
C=\frac{-\Lambda(4+\eta^2)}{4}\left(\frac{q^2\bar X_1+q\omega \bar X_2}{q^2-i\omega\frac{12-\eta^2}{4+\eta^2}}\right).
\end{equation}
From this expression, we ensure that the pole of the green's functions from such modes (odd parity modes) will be given by
\begin{equation}
\omega=-i\mathcal D q^2{\rm\  and \ \ }\mathcal D=\frac{4+\eta^2}{12-\eta^2}=\frac{-\Lambda(4+\eta^2)^2}{16\pi T_H(12-\eta^2)}
\end{equation}
where $\mathcal D$ is the diffusion constant. Diffusion constant is usually given by $\mathcal D=\tilde \eta/(\epsilon+P)$, where $\tilde \eta$ is the shear viscosity, $\epsilon$ is the energy density and
$P$ is the pressure of the dual fluid system. For the fluids whose gravity dual is long wavelength excitation in $AdS_{d+1}$ spacetime, $\mathcal D=\frac{1}{4\pi T_H}$
\cite{Kovtun:2003wp} (See the table.1 in this reference). By using the definition of the cosmological constant $\Lambda$ in (\ref{lambdaaaaa}), one realizes that $\mathcal D=\frac{1}{4\pi T_H}$
for our case too.

In fact, the relevant parts of the bulk on-shell action to obtain the retarded Green's functions in this case is given by
\begin{equation}
S_{bulk-os}=\frac{r^{2a_1}_h}{32\pi}\left(\frac{-\Lambda (4+\eta^2)}{4} \right)\int d\omega d^2q \left[ \bar X_2\left( \frac{\omega q \bar X_1+ \omega^2 \bar X_2}{q^2-i\omega\frac{12-\eta^2}{4+\eta^2}} \right)
+\bar X_1 \left( \frac{q^2 \bar X_1+ q\omega \bar X_2}{q^2-i\omega\frac{12-\eta^2}{4+\eta^2}}\right) + O(1/u)\right],
\end{equation}
and from this expression, the retarded  Green's functions can be read off as
\begin{eqnarray}
G^R_{ty,ty}&=&\frac{(4\pi T_H)^{\frac{8}{4-\eta^2}}}{16\pi}\left(\frac{4+\eta^2}{12-\eta^2}\right)^{\frac{4+\eta^2}{4-\eta^2}}
\frac{q^2 }{q^2-i\omega\frac{12-\eta^2}{4+\eta^2}}, \\
G^R_{xy,xy}&=&\frac{(4\pi T_H)^{\frac{8}{4-\eta^2}}}{16\pi}\left(\frac{4+\eta^2}{12-\eta^2}\right)^{\frac{4+\eta^2}{4-\eta^2}}
\frac{\omega^2 }{q^2-i\omega\frac{12-\eta^2}{4+\eta^2}}, \\
G^R_{ty,xy}&=&G^R_{xy,ty}=\frac{(4\pi T_H)^{\frac{8}{4-\eta^2}}}{16\pi}\left(\frac{4+\eta^2}{12-\eta^2}\right)^{\frac{4+\eta^2}{4-\eta^2}}
\frac{q\omega}{q^2-i\omega\frac{12-\eta^2}{4+\eta^2}}.
\end{eqnarray}

 \section{Discussion}
 \label{discussion}
In this note, we have discussed at length the transport coefficients of the non-conformal fluid system, whose gravity dual is Einstein-dilaton theory with a Liouville 
type dilaton potential parameterized by an intrinsic constant $\eta$. The ratio of the shear viscosity to entropy density in zero frequency limit turns
out to saturate the universal bound. However, the other 
transport coefficients in the hydrodynamic limit such as AC conductivity of the $R$-charge current, 
diffusion constant of the shear modes of the non-conformal fluids depend on $\eta$ non-trivially. 

We also computed the retarded Green's functions of R-charge current in the high frequency limit. 
It turns out that the retarded Green's functions
show exponentially decaying behaviors in momentum space as $G^R \sim e^{-\lambda |q|/T}$, where the constant $\lambda$ depends on $\eta$ in a non-trivial way.

For our next project, we will consider higher order conductivities from bulk $U(1)$ gauge fields in both frequency and spatial momenta and examine their dependence on $\et$.
We will also try to apply the WKB approximation scheme to other bulk perturbations in the high momenta limit and check its decaying behaviors. So far the interpretation of the retarded Green's functions in the high momenta limit is not clear in the dual fluid dynamics.
We would like to shed some light on this high momenta regime by studying the Einstein-dilaton system in gory details.

\vspace{1cm}

{\bf Acknowledgement}

This work was supported by the Korea Science and Engineering Foundation
(KOSEF) grant funded by the Korea government(MEST) through the Center for
Quantum Spacetime(CQUeST) of Sogang University with grant number R11-2005-021.
C. Park was also
supported by Basic Science Research Program through the
National Research Foundation of Korea(NRF) funded by the Ministry of
Education, Science and Technology(2010-0022369).

J.H.O would like thank CQUeST for the hospitality during his visit when a part of this work was accomplished. He also thanks his $\mathcal W.J.$
\vspace{1cm}


\end{document}